\documentclass[prl,twocolumn,nofootinbib,preprintnumbers,superscriptaddress]{revtex4-1}

\usepackage{natbib}
\usepackage{amssymb,amsbsy,amsmath,amsfonts}
\usepackage{mathrsfs}
\usepackage{graphicx}
\usepackage{epsf,epsfig,float,latexsym,amsthm,fancyhdr,rotating}
\usepackage{graphics,psfrag,longtable}
\usepackage{slashed}
\usepackage{esint}
\usepackage{nicefrac}
\usepackage{braket}
\usepackage{mathtools}

\newcommand{\nn}{\nonumber}
\renewcommand{\Im}{{\rm Im\,}} 
\renewcommand{\Re}{{\rm Re\,}}

\usepackage{mathptmx}   

\usepackage[colorlinks,citecolor=blue,linktoc=all,linkcolor=cyan]{hyperref} 

\DeclareMathAlphabet{\mathantt}{OML}{antt}{l}{it}
\DeclareMathAlphabet{\mathpzc}{OT1}{pzc}{m}{n}

\def\beq{\begin{equation}}
\def\eeq{\end{equation}}
\def\bea{\begin{eqnarray}}
\def\eea{\end{eqnarray}}
\def\beqa{\begin{equation}\begin{array}{l}}
\def\eeqa{\end{array}\end{equation}}

\def\barr{\left(\begin{array}{c}}
\def\earr{\end{array}\right)}
\def\bmat{\left(\begin{array}{cc}}
\def\emat{\end{array}\right)}

\begin{document}
\title {Dilepton photoproduction on a deuteron target}

\author{Carl E. Carlson}

\affiliation{College of William and Mary, Physics Department, Williamsburg, Virginia 23187, USA}

\author{Vladyslav Pauk}

\affiliation{Institut f\"ur Kernphysik, Cluster of Excellence PRISMA,  Johannes Gutenberg-Universit\"at, 
Mainz, Germany}

\author{Marc Vanderhaeghen}

\affiliation{Institut f\"ur Kernphysik, Cluster of Excellence PRISMA,  Johannes Gutenberg-Universit\"at, 
Mainz, Germany}

\begin{abstract}
We investigate the sensitivity of the cross section for lepton pair production off a deuteron target, $\gamma d \to l^+ l^- d$,  to the deuteron charge radius.  We show that for small momentum transfers the Bethe-Heitler process dominates, and that it is sensitive to the charge radius such that a cross section ratio measurement of about $0.1 \%$ relative accuracy could give a deuteron charge radius more accurate that the current electron scattering value and sufficiently accurate to distinguish between the electronic and muonic atomic values.
\end{abstract}
\date{\today}
\maketitle

Over the past decade, the extractions of the proton charge radius from the Lamb shift measurements in muonic hydrogen \cite{Pohl:2010zza,Antognini:1900ns} resulted in a significant discrepancy in comparison with measurements with electrons \cite{Bernauer:2010wm,Bernauer:2013tpr,Mohr:2015ccw}, amounting to a $5.6~\sigma$ difference according to a recent re-evaluation~\cite{Krauth:2017ijq}. 
The resolution of this ``proton radius puzzle" has triggered a lot of activity, 
see e.g. Refs.~\cite{Pohl:2013yb,Carlson:2015jba,Hill:2017wzi} for recent reviews. 
Corresponding measurements on the deuteron have not only confirmed the puzzle~\cite{Pohl1:2016xoo}, but have also revealed a $3.5~\sigma$ difference between the spectroscopic measurements in muonic versus ordinary deuterium. The deuteron charge radius as extracted from elastic electron scattering~\cite{Sick:1998cvq} has at present a too large error bar to distinguish between both spectroscopic values. 
In this letter, we investigate the sensitivity of the complementary lepton pair production process off a deuteron target, $\gamma d \to l^+ l^- d$,  to the deuteron charge radius.

\begin{figure}[t]
\centering
\includegraphics[width=0.75\columnwidth]{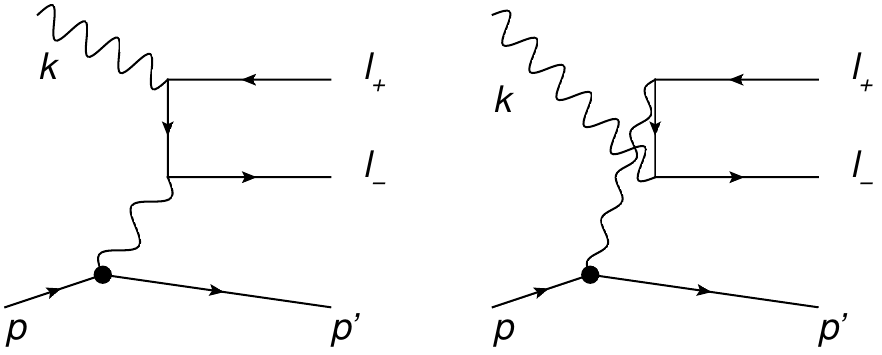}
\vskip 2 mm
\centerline{   \includegraphics[width=0.46\columnwidth]{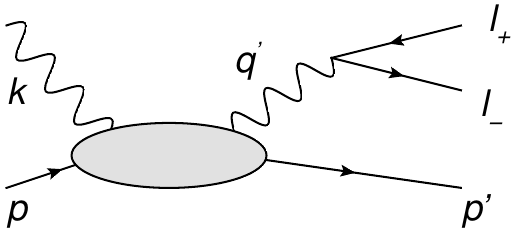}   }
\caption{Mechanisms for $\gamma d \to l^-l^+ d$. The momenta of the external particles are $k$ for the photon, $p (p^\prime)$ for initial (final) deuterons, and $l_-$, $l_+$ for the lepton pair.  The upper diagrams show the Bethe-Heitler mechanism; the lower diagram shows the Compton mechanism.} 
\label{fig1}
\end{figure}

We consider 
$\gamma d \to l^- l^+ d$ in the limit of 
very small spacelike momentum transfer, defined as $\Delta \equiv p^\prime - p$, with 
four-momenta as indicated on Fig.~\ref{fig1}. Furthermore, we will use in the following the Mandelstam invariant $s = (k + p)^2 = M_d^2 + 2 M_d E_\gamma$, with $M_d$ the deuteron mass and $E_\gamma$ the photon {\it lab} energy, 
the Mandelstam invariant $t = \Delta^2$, 
as well as the squared invariant mass of the lepton pair, defined as 
$M_{ll}^2 \equiv (l_- + l_+)^2$. 
In the limit of small $-t$, the Bethe-Heitler (BH) mechanism, shown in Fig.~\ref{fig1} dominates the cross section of the $\gamma d \to l^- l^+ d$ reaction, as we shall show.

The deuteron electromagnetic structure entering the hadronic vertex in the BH process of Fig.~\ref{fig1} is described by three elastic electromagnetic form factors (FFs), 
corresponding to the Coulomb monopole ($G_C$), magnetic dipole ($G_M$), 
and Coulomb quadrupole ($G_Q$) FFs, respectively.  
The definitions and normalizations of $G_C$, $G_M$, and $G_Q$ follow from~\cite{Arnold:1979cg},
\begin{align}
&\left\langle p^\prime,\lambda^\prime \right|  J^\mu (0) \left| p, \lambda \right\rangle 
= \varepsilon_\alpha (p, \lambda) \, \varepsilon_\beta^\ast (p^\prime, \lambda^\prime) 
	\Bigl\{ 
-  g^{\alpha \beta} \, 2P^\mu G_1(t)	\nn\\		
& \hskip 2.5 em - \left( g^{\alpha \mu}  \Delta^\beta - g^{\beta \mu}  \Delta^\alpha \right) G_M(t) 
+ \Delta^\alpha  \Delta^\beta \, \frac{P^\mu}{M_d^2} G_3(t)
		\Bigr\},   
\label{eq:emcurrent}
\end{align}
where $P=(p+p')/2$, and $\varepsilon_\alpha$ and $\varepsilon_\beta^*$ are deuteron polarization vectors.   The charge and quadrupole FFs follow from,
\begin{eqnarray}
G_C = G_1 + \frac{2}{3} \tau_d \, G_Q, 	
\quad 
G_Q = G_1 - G_M + \left(1+ \tau_d \right) G_3,
\end{eqnarray}
with normalizations $G_C(0)=1$, 
$G_M(0) = \mu_d$ (magnetic moment in units $e/(2M_d)$), 
$G_Q(0) = Q_d$ (quadrupole moment in units $e/M_d^{2}$), 
and where $\tau_d \equiv -t / (4 M_d^2)$. 
For numerical evaluation, we will use the parameterization of the deuteron FFs, obtained from scattering and tensor polarization data, and given as fit II by Abbott et al.~\cite{Abbott:2000ak}; see also~\cite{Carlson:2008zc} for details.  In addition, we have the deuteron FF parameterization based on scattering data, but including a treatment of two and more photon exchange corrections, by Sick and Trautmann~\cite{Sick:1998cvq}.

As the momentum transfer $t$ is the argument appearing in the form factor (FF) in the BH process, a measurement of the cross section in the small $-t$ kinematic regime, where the BH process dominates, will allow accessing the deuteron charge FF $G_{C}$ at small spacelike momentum transfer.   The deuteron charge radius $R_d$ is
determined from $G_C$  through
\begin{eqnarray}
G_C(t) = 1 + \frac{1}{6} R_d^2 \, t \, + \mathcal{O}(t^2)	.
\label{eq:taylorgc}
\end{eqnarray} 

We quote several current values for the deuteron charge radius $R_d$, all in femtometers.  

\begin{align}
R_d = 	\left\{	\begin{array}{ll}
	2.088 
				& \text{Abbott et al.~fit~\cite{Abbott:2000ak}}	,			\\[0.2ex]
	2.130(10) 
				& e\text{-}d \text{ elastic scattering~\cite{Sick:1998cvq}}	,	\\[0.2ex]
	2.1415(45) 
				& \text{atomic deuterium spectroscopy~\cite{Pohl:2016glp}},	\\[0.2ex]
	2.1413(25) 
				& \text{CODATA 2014~\cite{Mohr:2015ccw}},			\\[0.2ex]
	2.12562(78) 
				& \mu\text{-}d \text{ Lamb shift~\cite{Pohl1:2016xoo}},		\\[0.2ex]
	2.12771(22) 
				& \mu\text{-}H  \text{ Lamb shift \& isotope shift~\cite{Pohl1:2016xoo}}.
			\end{array}	\right.
\end{align}
Of the two purely or mainly atomic values,  atomic deuteron spectroscopy uses only fits to energy splittings measured in deuterium, while CODATA uses the proton radius obtained electronically and the isotope shift (the very accurate measurement of $R_d^2 - R_p^2$~\cite{Parthey:2010aya}, using ordinary hydrogen).  The last listed radius measurement also uses the isotope shift, this time combined, supposing the absence of new physics, with the proton radius measured from the muonic hydrogen Lamb shift.  Notable is the definite incompatibility between the deuteron radius measured using ordinary and muonic atoms.

\begin{figure}[]
\centering
\includegraphics[width=\columnwidth]{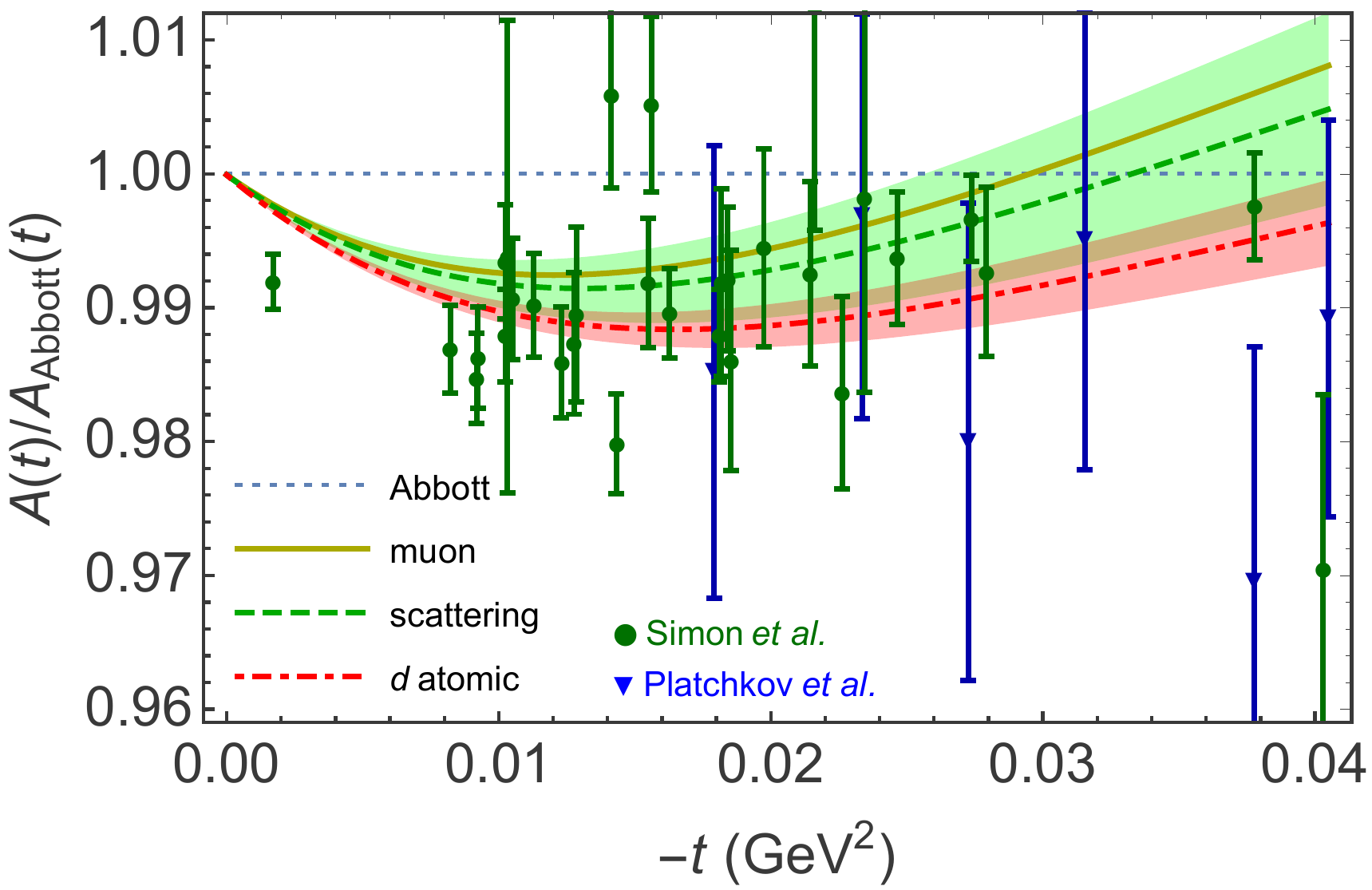}

\caption{Three predicted results for $ed$ elastic scattering normalized to the Abbott 
\textit{et al.}~parameterization~\cite{Abbott:2000ak}, with data from Simon \textit{et al.}~\cite{Simon:1981br} and Platchkov \textit{et al.}~\cite{Platchkov:1989ch}.  The deuteron charge radii are from the muonic deuterium Lamb shift~\cite{Pohl1:2016xoo} (gold solid line, with uncertainty comparable to the width of the line);   from $e$-$d$ elastic scattering~\cite{Sick:1998cvq} (green dashed line, with uncertainty limits indicated by the green band);   from deuterium atomic spectroscopy~\cite{Pohl:2016glp} (red dot-dashed line, with uncertainty limits indicated by the red band).  The CODATA deuteron radius~\cite{Mohr:2015ccw} would be identical, on this scale, to the red dot-dashed line but with an uncertainty band \nicefrac{5}{9} as wide.  The data were given the McKinley-Feshbach~\cite{McKinley:1948zz} two photon corrections.
}
\label{fig:elasticrelative}
\end{figure}

As a preliminary observation, the effect of the radius modifications on the calculated $e$-$d$ elastic scattering cross section is shown in Fig.~\ref{fig:elasticrelative}, where cross sections are shown relative to the Abbott {et al.} results. Results using the Sick-Trautmann parameterization are labeled scattering, and results obtained for other values of $R_d$ are obtained by modifying the Sick-Trautmann  $G_C$ form factor as,
\begin{eqnarray}
G_C(t) = \frac{G_{C, \, \text{Sick-Trautmann}}(t)}
	{\left[1 - \frac{1}{6} (R_d^2 - R_{d,\, \text{Sick-Trautmann}}^2) \, t \right]}.
\label{eq:gccorr}
\end{eqnarray}
This will allow studying the dependence on $R_d$ while keeping the same curvature terms as used in the Sick-Trautmann parameterization of the $G_C$ FF.

The elastic cross sections are obtained from the no structure cross section and the form factors as~\cite{Arnold:1979cg}
\begin{align}
\frac{d\sigma}{d\Omega} = \left.  \frac{d\sigma}{d\Omega}  \right|_\text{NS}
	\left[ A(t) + B(t) \tan^2 (\theta_e/2) \right]		,
\end{align}
with $\theta_e$ the electron lab scattering angle, and where
\begin{align}
A(t) &= G_{C}^2(t) + \frac{2}{3} \tau_d G_{M}^2(t) 
	+ \frac{8}{9} \tau_d^2 G_{Q}^2(t),		\nonumber\\
B(t) &= \frac{4}{3} \tau_d \left( 1 + \tau_d \right)G_{M}^2(t)  .
\end{align}

Experimental data for Simon {et al.}~\cite{Simon:1981br} and Platchkov {et al.}~\cite{Platchkov:1989ch} are also shown.  The two photon corrections as given by McKinley and Feshbach~\cite{McKinley:1948zz} have been applied to the data.

Turning to leptoproduction,  the differential cross section for the BH process is strongly peaked for leptons emitted in the incoming photon direction, and as we aim to maximize the BH contribution in this work in order to access $G_{C}$, 
we will study the $\gamma d \to l^- l^+ d$  process when (only) detecting the recoiling deuteron's momentum and angle, thus effectively integrating over the large lepton peak regions.  
The {\it lab} momentum of the deuteron is in one-to-one relation with the momentum transfer $t$: 
$|\vec p^{\, \prime}|^{lab} = 2 M_d \sqrt{\tau_d (1 + \tau_d)}$. Furthermore, for a fixed value of $t$, 
the recoiling deuteron {\it lab} 
angle $\Theta_d^{lab}$ is expressed in terms of invariants as~:
\begin{eqnarray}
\cos \Theta_d^{lab} = \frac{M_{ll}^2 + 2 (s + M_d^2) \tau_d}{2 (s - M_d^2) 
\sqrt{\tau_d (1 + \tau_d)}}.
\end{eqnarray}

The differential cross section for the dominating BH process to 
the $\gamma d \to l^- l^+ d$ reaction, differential in $t$, $M_{ll}^2$, and the lepton solid angle 
$\Omega^{l^-l^+ \text{cm}}$ in the c.m.~frame of the dilepton pair is given by
\begin{eqnarray}
\label{eq:sigmaelastic}
\frac{d \sigma^{BH}}{dt \, dM_{ll}^2 \, d \Omega^{l^- l^+ \text{cm}}} &=& 
\frac{\alpha^3 \beta}{16 \pi (s - M_d^2)^2 \, t^2}  L_{\mu \nu} H^{\mu \nu},
\label{eq:cross1}
\end{eqnarray}
with $\alpha \equiv e^2 / 4 \pi \approx 1/137$, 
and $\beta \equiv \sqrt{1 - {4 m^2}/{M_{ll}^2}}$ 
the lepton velocity in the $l^- l^+$ {\it c.m.} frame, with $m$ the lepton mass. 
Furthermore in Eq.~(\ref{eq:cross1}), $L_{\mu \nu}$ is the unpolarized lepton tensor, averaged over the initial photon polarizations, given by:
\begin{eqnarray}
L_{\mu \nu} &=&
 { - \frac{1}{2}} 	\mathrm{Tr}  \bigg\{  (\slashed{l}_- + m ) 
 	\left[ \gamma^\alpha \frac{\slashed{l}_- - \slashed{k} + m}{- 2 k \cdot l_- } \gamma_\mu 
+ \gamma_\mu \frac{\slashed{k} - \slashed{l}_+  + m}{- 2 k \cdot l_+ }
\gamma^\alpha \right] 
			\nonumber \\
&\times& 	(\slashed{l}_+ - m )  
	\left[ \gamma_\nu \frac{\slashed{l}_- - \slashed{k} + m}{- 2 k \cdot l_- } \gamma_\alpha 
+ \gamma_\alpha \frac{\slashed{k} - \slashed{l}_+  + m}{- 2 k \cdot l_+ }
\gamma_\nu \right]  
\bigg\},  
\end{eqnarray}
and $H^{\mu \nu}$ is the unpolarized hadronic tensor defined by:
\begin{eqnarray}
\label{eq:hadrontensor}
H^{\mu \nu} &=& \frac{1}{3} \sum_{\lambda = 0, \pm 1} \sum_{\lambda^\prime = 0, \pm 1} 
\left\langle p^\prime,\lambda^\prime \right|  J^\mu (0) \left| p, \lambda \right\rangle  \nonumber \\
&&\hspace{2.5cm} \times 
\left\langle p^\prime,\lambda^\prime \right|  J^\nu (0) \left| p, \lambda \right\rangle^\ast . 
\end{eqnarray}
Using Eq.~(\ref{eq:emcurrent}), the hadronic tensor for the BH process can be expressed as:
\begin{eqnarray}
H^{\mu \nu} &=& \left( - g^{\mu \nu} + \frac{\Delta^\mu \Delta^\nu}{\Delta^2} \right) 
\left[ \frac{8}{3} M_d^2 \tau_d (1 + \tau_d) \, G^2_M \right] \nonumber \\
&+& 4\, P^\mu P^\nu \left[ G^2_C + \frac{2}{3} \tau_d \, G^2_M + \frac{8}{9} \tau_d^2 \, G^2_Q \right],
\end{eqnarray}
and the FFs are functions of the momentum transfer $t$.   

\begin{figure}[t]
\centering
\includegraphics[width=\columnwidth]{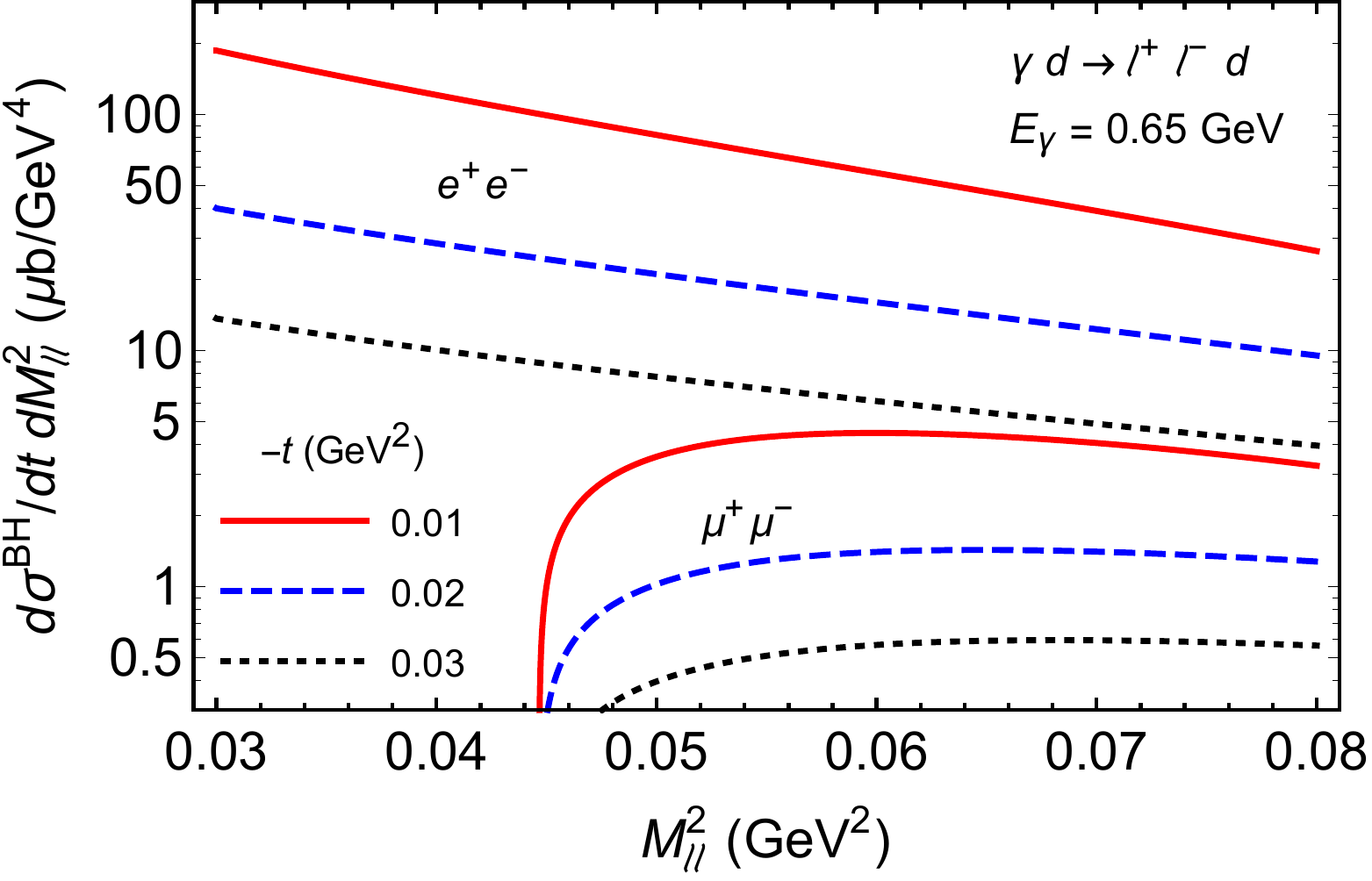}
\caption{Absolute cross sections, using the Abbott \textit{et al.}~\cite{Abbott:2000ak} deuteron FFs, showing the lepton pair invariant mass dependence of the $\gamma d \to e^+e^- d$ process (upper three curves) and the $\gamma d \to \mu^+ \mu^- d$ process (lower three curves) at $E_\gamma = 0.65$~GeV.  Three values of the momentum transfer are shown. } 
\label{fig:inelasticabsolute}
\end{figure}

When detecting only the deuteron momentum and angle,  the cross section 
integrated over the lepton angles is
\begin{eqnarray}
\frac{d \sigma^{BH}}{dt \, dM_{ll}^2} &=&  
\frac{ 4 \alpha^3   \beta}{(s - M_d^2)^2 \, t^2 (M_{ll}^2 - t)^4}  \nonumber \\
&\times& \left\{ C_E \left( G_C^2 + \frac{8}{9} \tau_d^2 G_Q^2 \right)   
+ C_M \frac{2}{3}  \tau_d  G_M^2 \right\},
\label{eq:intcross}
\end{eqnarray}
where
\begin{eqnarray}
C_{E, M} = C_{E, M}^{(1)} + C_{E, M}^{(2)} 
\frac{1}{\beta} \ln \left( \frac{1 + \beta}{1 - \beta} \right)	.
\label{eq:CEM}
\end{eqnarray}

The coefficients $C_{E, M}^{(1)}$, and $C_{E, M}^{(2)}$ are expressed through invariants as
\begin{align}
\label{eq:cem}
C_E^{(1)} &= t \left(s - M_d^2 \right) \left( s - M_d^2 - M_{ll}^2 + t \right) 
						\nonumber\\
&\qquad \times	\left[ M_{ll}^4 + 6 M_{ll}^2 t + t^2 + 4 m^2 M_{ll}^2 \right] 
						\nonumber \\
&+ \left(M_{ll}^2 - t \right)^2 
	\left[ t^2 M_{ll}^2 + M_d^2 (M_{ll}^2 + t )^2 + 4 m^2 M_d^2 M_{ll}^2 \right], \nn
						\\
C_E^{(2)} &= - t \left(s - M_d^2 \right) \left( s - M_d^2 - M_{ll}^2 + t \right) 
						\nonumber\\
&\qquad \times		\left[ M_{ll}^4 + t^2 + 4 m^2 \left( M_{ll}^2 + 2 t - 2 m^2\right) \right] 
						\nonumber \\
&+ \left(M_{ll}^2 - t \right)^2 
						\nonumber\\
&\hskip - 1.5 em	\times \big[ {-} M_d^2 (M_{ll}^4 + t^2 ) 
 		+ 2 m^2 \left( - t^2 {-} 2 M_d^2 M_{ll}^2 + 4 m^2 M_d^2 \right) \big],  \nn
						\\
C_M^{(1)} &= C_E^{(1)} - 2 M_d^2 (1 + \tau_d)   
						\left(M_{ll}^2 - t \right)^2  
\left[  M_{ll}^4 + t^2 + 4 m^2 M_{ll}^2 \right], 
						\nn\\
C_M^{(2)} &= C_E^{(2)} + 2 M_d^2 (1 + \tau_d)   
\left(M_{ll}^2 - t \right)^2  	
						\nonumber\\
&\qquad \times	\left[  M_{ll}^4 + t^2 + 4 m^2 \left( M_{ll}^2 - t - 2 m^2 \right) \right] .
\end{align}

The absolute Bethe-Heitler cross sections are shown in Fig.~\ref{fig:inelasticabsolute}. The abscissa is the dilepton mass-squared $M_{ll}^2$, with both electrons and muons represented, and showing three different values of $-t$.   The plot is similar to the one for protons~\cite{Pauk:2015oaa}, but the cross sections are smaller because of the faster falloff with $|t|$ of the deuteron FFs.

To estimate the Compton mechanism, the lower graph in Fig.~\ref{fig1}, we estimate the $S$-matrix amplitude
\begin{align}
&\mathcal M_C = - \frac{ e^3 }{ {q'}^2 } \epsilon_\nu(k,\lambda_\gamma) 
	\bar u(l_-,s_-) \gamma_\mu v(l_+,s_+)		\nn\\
&\times	\int d^4x\,	e^{iq'x}	\braket{ p',\lambda' | T J^\mu(x) J^\nu(0) | p,\lambda }	\\ \nn
&	\equiv	-i \frac{ e }{ {q'}^2 } \epsilon_\nu(k,\lambda_\gamma) 
	\bar u(l_-,s_-) \gamma_\mu v(l_+,s_+)	\,	8\pi M_d T_\text{TCS}^{\mu\nu}(k,q',P)   ,
\end{align}
where $T^{\mu\nu}_\text{TCS}$ is the unpolarized timelike real Compton tensor and $\lambda_\gamma$ is the photon polarization.

\begin{figure}[t]
\centering
\includegraphics[width= 1.00 \columnwidth]{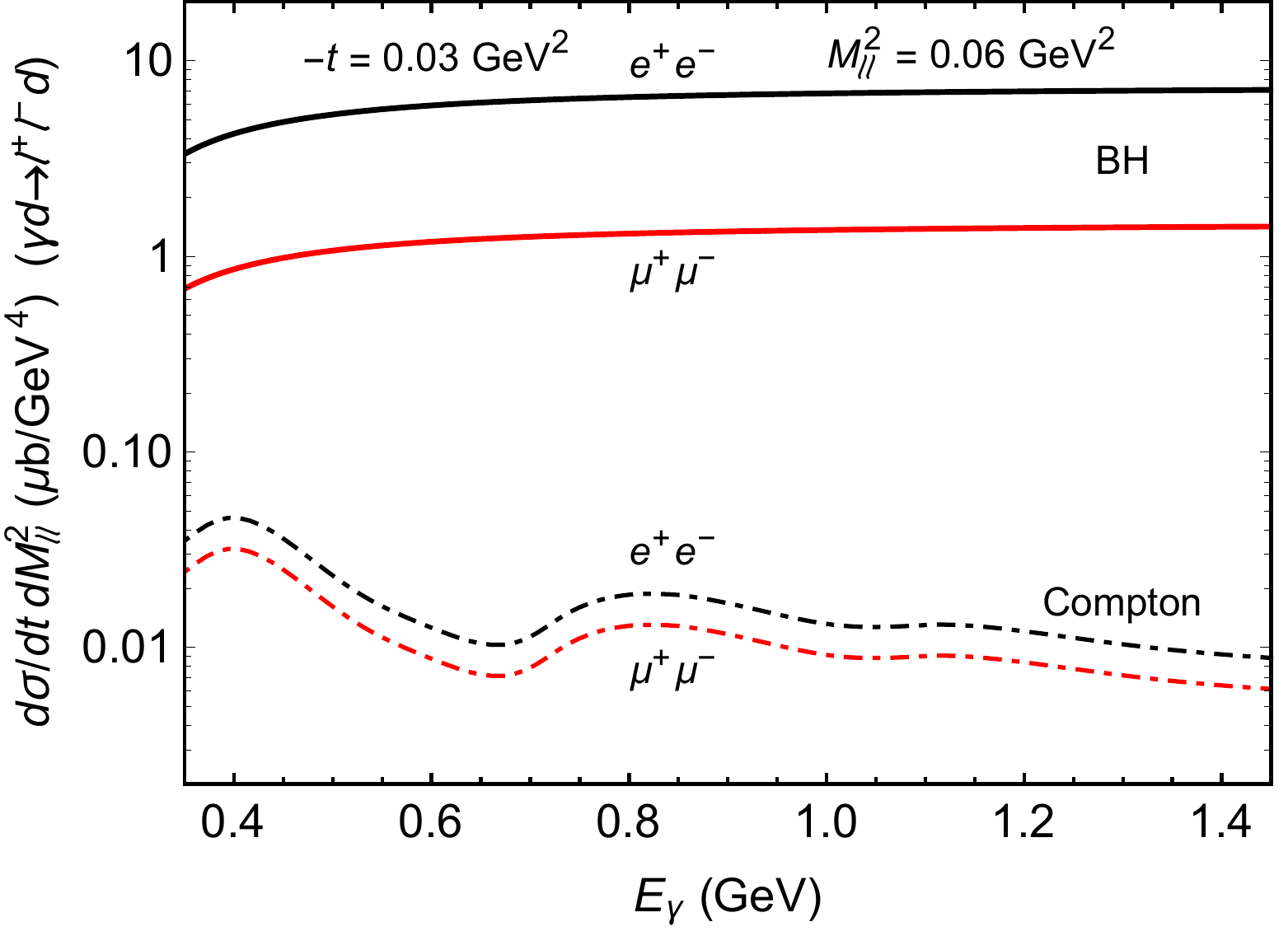}
\caption{The differential cross section for $\gamma d \to l^+ l^- d$ plotted vs. $E_\gamma$ and showing both the Bethe-Heitler and Compton contributions, for values of $t$ and dilepton mass indicated.} 
\label{fig:inelasticabsolute2}
\end{figure}

For near-real, near-forward kinematics, $M_{ll}^2, |t| \ll s$ the unpolarized TCS amplitude can be approximated by,
\begin{align}
T^{\mu\nu}_\text{TCS}(k,q',P) \approx 
	\left(g^{\mu\nu} - \frac{ {q'}^\mu k^\nu}{q' \cdot k } \right)	T_{1d}(\nu,t,M_{ll}^2)	,
\end{align}
where $T_{1d}$ denotes the leading scalar amplitude, and $\nu$ the crossing symmetric variable, defined as

For $M_{ll}^2, |t| \ll s$, we can further approximate
\begin{align}
T_{1d}(\nu,t,{q'}^2) \approx f(\nu) 	,
\end{align}
where $f(\nu)$ is the unpolarized forward real Compton amplitude for a deuteron target.
Its imaginary part can be obtained from the photoproduction total cross section, or from the $F_{1d}(\nu,Q^2)$ structure function,  as
\begin{align}
\Im f(\nu) = \frac{ \nu }{ 4\pi } \sigma(\nu) = \frac{ \pi \alpha }{ M_d } F_{1d}(\nu,0)		.
\end{align}

Analyticity and the low-energy theorem value of $f(0)$ allow us to  obtain the real part of $f$ from a once-subtracted dispersion relation,
\begin{align}
\Re f(\nu) = - \frac{ \alpha }{ M_d } + \frac{\nu^2 }{ 2\pi^2 } \fint_{\nu_0}^\infty
	d\nu' \frac{ \sigma(\nu') }{ {\nu'}^2 - \nu^2 }	,
\end{align}
where $\nu_0$ is the inelastic threshold, 
$\nu_0 = ((M_n+M_p)^2-M_d^2)/(2M_d) 
\approx 2.23$ MeV.

The Compton contribution to the $\gamma d \to l^+ l^- d$ differential cross section, integrated over lepton angles, is
\begin{align}
\frac{ d\sigma^\text{TCS} }{ dt \, dM_{ll}^2 } 
	= \frac{ 2 M_d^2 \alpha^3 \beta  } { (s-M_d^2)^2  M_{ll}^2  }
		\left(1 - \frac{ \beta^2 }{ 3 } \right) \left| \frac{ f(\nu) }{ \alpha } \right|^2	.
\end{align}

Figure~\ref{fig:inelasticabsolute2} shows the Compton cross section, compared to the Bethe-Heitler, for particular values of $t$ and $M_{ll}^2$, with $E_\gamma$ on the abscissa.  We obtained $\sigma(\nu)$, or $F_{1d}(\nu,0)$ in the quasi-elastic region from the fits of~\cite{Carlson:2013xea}, and in the nucleon inelastic region from Bosted-Christy~\cite{Bosted:2007xd} deuteron fits when ${s_\text{nucleon}} < (3.1 $ GeV)$^2$ and from Capella et al.~\cite{Capella:1994cr}, isospin modified for the neutron, above that.  
There is no interference between the Compton and Bethe-Heitler contributions when we integrate over the lepton angles.  The Compton cross section is more than two orders of magnitude smaller than the Bethe-Heitler cross section, for this energy and momentum transfer range.

\begin{figure}[t]
\centering
\includegraphics[width=\columnwidth]{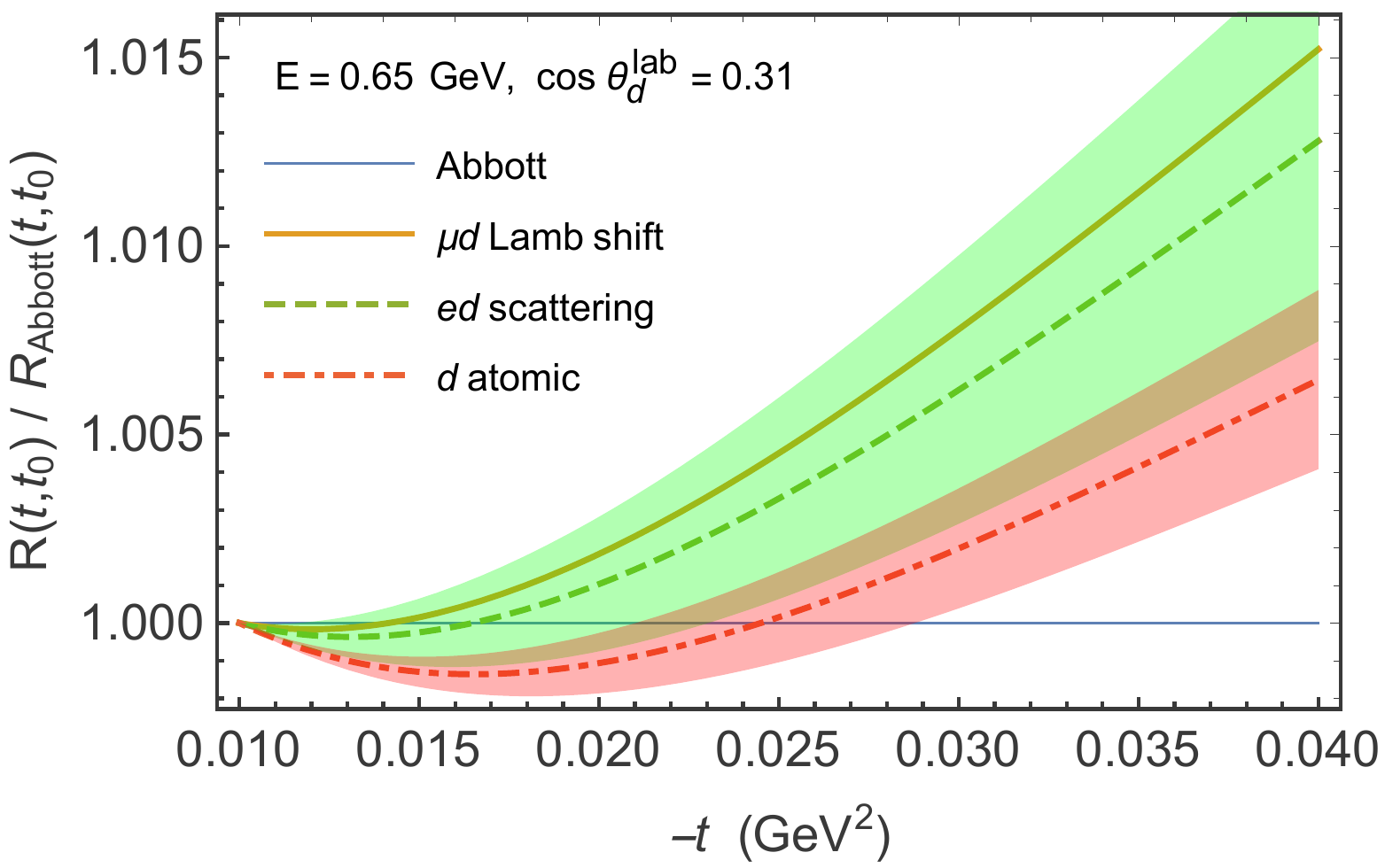}
 \vglue - 1 mm
\caption{The  $t$-dependence of the $\gamma d \to e^+e^- d$ cross section,  $R(t,t_0) \equiv  d\sigma / dt \, dM_{ll}^2(t) /  d\sigma / dt \, dM_{ll}^2(t_0)$,  relative to a reference value $t_0 = -0.01$ GeV$^2$, at fixed outgoing deuteron lab angle, for beam energy 0.65 GeV.  The ratio is normalized to the result for Abbott et al.~FFs~\cite{Abbott:2000ak}.  The different deuteron radii and associated error bands are as in Fig.~\ref{fig:elasticrelative}. 
}
\label{fig:inelasticrelative}
\end{figure}

The sensitivity of the differential inelastic cross section for different FFs and different deuteron radii is shown in Fig.~\ref{fig:inelasticrelative}, as a function of $- t$, with a photon beam energy  0.65 GeV, corresponding with a minimum in the Compton contribution.  The plot shows the cross section relative to a reference value $t_0 = -0.01$ GeV$^2$,  and normalized to results from the Abbott {et al.}~parameterization.  The outgoing deuteron angle has been fixed.  The fixed angle can allow better experimental calibration than in an elastic scattering experiment, where different momentum transfers require different scattering angles, for a given beam energy.    Measurements of such a fixed angle ratio at different $t$ with $0.1\%$ relative accuracy would allow distinguishing the various fits.

In this work we have studied dilepton photoproduction off a deuteron with the aim of extracting the deuteron charge radius.
By studying the momentum transfer dependence of the outgoing deuteron at a fixed angle, we have seen that a cross section ratio measurement of about 0.1\% accuracy will allow extracting a deuteron charge radius more accurate than the present value from elastic scattering, and can distinguish between the values obtained from ordinary and muonic deuterium, which are currently at variance by around $3.5~\sigma$. 

\section*{Acknowledgements}
We thank Keith Griffioen and Tim Hayward for useful conversations, and Ingo Sick for providing details of 
the fits in~\cite{Sick:1998cvq}.  CEC thanks the National Science Foundation (USA) for support under grant PHY-1516509, and the Johannes Gutenberg-University, Mainz, for hospitality while this work was completed.  The work of VP and MV was supported by the Deutsche Forschungsgemeinschaft DFG in part through the Collaborative Research Center [The Low-Energy Frontier of the Standard Model (SFB 1044)], and in part through the Cluster of Excellence [Precision Physics, Fundamental Interactions and Structure of Matter (PRISMA)].

\bibliography{dilepton}

\end{document}